\documentclass{article}

\usepackage{arxiv}

\usepackage{geometry}               % Page layout
\usepackage{multicol}               % Multiple columns
\usepackage{enumitem}               % Customizable lists
\usepackage{caption}                % Customizing captions
\usepackage{xcolor}                 % Helper functions for colors
\usepackage{float}

\usepackage{fontspec}               % Font selection for LuaLaTeX
\usepackage{microtype}              % Microtypography improvements
\usepackage{unicode-math}           % Unicode math support
\usepackage{amsmath}                % AMS math environments and commands
\usepackage{amsthm}                 % Theorem environments
\usepackage{amsfonts}               % Mathematical symbols
\usepackage{nicefrac}               % Compact symbols for fractions
\usepackage{braket}                 % Nice <bra|ket> notation
\usepackage{siunitx}
\usepackage{booktabs}               % Professional quality tables
\usepackage{array}                  % Extended tabular environments
\usepackage{graphicx}               % Including graphics
\usepackage{tabularx}
\newcolumntype{Y}{>{\raggedright\arraybackslash}X}
\usepackage{caption}                % Caption for figures in multicol
\usepackage{listing}
\usepackage{multirow}

\usepackage{xcolor}                 % Color support
\definecolor{fadedrose}{HTML}{a46167}
\definecolor{salmon}{HTML}{ea9971}
\definecolor{darkblue}{HTML}{525b7a}
\definecolor{darkersalmon}{HTML}{da6d5d}
\definecolor{navyblue}{HTML}{183d61}
\definecolor{taupe}{HTML}{e6c197}
\definecolor{fadedpurple}{HTML}{785f72}

\definecolor{brightochre}{HTML}{f1ab57}
\definecolor{brightsalmon}{HTML}{f17152}
\definecolor{brightcherry}{HTML}{c63f5f}
\definecolor{brightnavy}{HTML}{284a8b}

\usepackage{hyperref, url}          % Hyperlinks and URLs
\hypersetup{
    colorlinks=true,                % Enable colored text instead of boxes
    linkcolor=brightsalmon,         % Color for internal links
    urlcolor=brightsalmon,          % Color for URLs
    citecolor=brightsalmon,         % Color for citations
    filecolor=brightsalmon          % Color for file links
}
\usepackage{orcidlink}              % ORCID links

\usepackage[
  backend=biber,
  style=numeric-comp,
  sorting=ynt,
]{biblatex}
\AtEveryBibitem{%
  \iffieldundef{doi}{}{\clearfield{eprint}}%
}
\AtEveryBibitem{%
  \clearfield{issn}%
}
\AtEveryBibitem{%
  \clearfield{url}%
}

\usepackage{lipsum}                 % Lorem ipsum text for testing
\usepackage{printlen}
\usepackage{csquotes}
\usepackage{placeins} % don't let floats to float beyong section borders
\newcommand{\emoji}[1]{\raisebox{-0.2\height}{\includegraphics[height=1em]{figures/emojis/#1.png}}}

\newcommand{\twocolumngeometry}{
    \newgeometry{
        left=1.55cm,
        right=1.55cm,
        columnsep=0.8cm,
        top=3cm,
        bottom=3cm,
    }
}

\newenvironment{centeredfigure}
{\begin{center} %
}{\end{center} %
}

\newcommand{\captionbf}[1]{\textbf{#1}}

\usepackage{algorithmic}
\usepackage[ruled]{algorithm2e}

\SetCommentSty{mahagonyComment}
\SetAlFnt{\small}
\SetAlCapNameFnt{\small}
\SetAlCapFnt{\small}
\SetAlCapSty{}
\newcommand{\openwidetext}{\end{multicols}\rule{\dimexpr(0.5\textwidth-0.5\columnsep-0.4pt)}{0.4pt}\rule{0.4pt}{6pt}}
\newcommand{\closewidetext}{\hfill\rule[-6pt]{0.4pt}{6.4pt}\rule{\dimexpr(0.5\textwidth-0.5\columnsep-1pt)}{0.4pt}\begin{multicols}{2}}



\newcommand{\cyrillicdivider}{\raisebox{-0.5\height}{\includegraphics[height=1em]{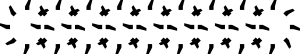}}}

\title{TeraGram: A Structured Longitudinal Dataset of the Telegram Messenger}

\usepackage{authblk}

\author[1, 2, $\Omega$]{Anastasia Golovin~\orcidlink{0009-0005-3490-1354}}
\author[1, 2, $\Omega$]{Sebastian B. Mohr~\orcidlink{0000-0002-4721-0561}}
\author[3, 1, 4]{Arne I. Gottwald~\orcidlink{0009-0002-0161-4453}}
\author[5, 6]{Ulrik Hvid~\orcidlink{0009-0009-5863-2012}}
\author[7, 8]{Srushhti Trivedi~\orcidlink{0009-0000-6584-4448}}
\author[9]{Joao P. Neto~\orcidlink{0000-0003-3027-8276}}
\author[1, 2]{Andreas C. Schneider}
\author[1, 2]{Viola Priesemann}

\affil[1]{Max Planck Institute for Dynamics and Self-Organization, Göttingen, Germany.}
\affil[2]{Institute for the Dynamics of Complex Systems, University of Göttingen, Göttingen, Germany.}
\affil[3]{Campus Institute for Dynamics of Biological Networks, University of Göttingen, Göttingen, Germany.}
\affil[4]{Campus Institute Data Science, University of Göttingen, Göttingen, Germany.}
\affil[5]{Biocomplexity, Niels Bohr Institute, University of Copenhagen, Copenhagen, Denmark}
\affil[6]{PandemiX – Center for Interdisciplinary Study of Pandemic Signatures, Copenhagen, Denmark}
\affil[7]{Institute of Medical Informatics, University Medical Center Göttingen, Germany}
\affil[8]{Institute for Ethics and History of Medicine, University Medical Center Göttingen, Germany}
\affil[9]{IDea\_Lab, University of Graz, Graz, Austria}
\affil[$\Omega$]{These authors contributed equally.}

\begin{document}
\maketitle

\begin{abstract}
Here we present a massive longitudinal dataset of public Telegram content, comprising over 5.9 billion messages dating from 2015 to 2025, collected from 712 thousand channels and groups, enriched with metadata on forwards, reactions, and polls. The dataset spans multiple languages including Russian and Farsi, representing countries where Telegram shows mainstream adoption, as well as Western languages where Telegram is used in specific sub-communities. 
The dataset has several advantages. First, when restricted by language, it provides a versatile example of an algorithm-free platform, contrary to many other social media platforms that are strongly influenced by opaque content-curation algorithms. Second, it enables comparative studies across different languages, communities, and user bases under identical platform affordances. 
The dataset thus offers a foundation for studying engagement patterns, network evolution, and community formation in the absence of algorithmic curation.

\end{abstract}

\twocolumngeometry

\begin{multicols}{2}

\section{Introduction}

Social media platforms play a central role in shaping public opinion \cite{mcgregor2019social,okechukwu2023media}. However, researchers seeking to understand this influence face a persistent challenge: most platforms use sophisticated and often proprietary algorithms to filter and prioritize content. When algorithms determine what users see, it is challenging to disentangle organic social dynamics from effects induced by proprietary algorithms.

Telegram stands out as a platform with minimal algorithmic intervention. Originally launched as an instant messaging app, it has evolved into a hybrid platform that combines private messaging with large-scale public broadcasting. Unlike most mainstream platforms, Telegram features a chronological feed and little to no algorithmic recommendations. Its emphasis on privacy and limited moderation has attracted activists and journalists who are seeking secure communication in authoritarian regimes \cite{SuTelegramAntiELAB2022, WijermarsTelegramHarbinger2022, UrmanAnalyzingProtest2021}; it also made Telegram a hub for misinformation, extremist content, and illicit activities \cite{KiessEuroscepticismLocal2025, WaltherUSExtremism2021}. This has led to pronounced differences in adoption and public sentiment towards the platform across countries: Telegram is widely adopted and trusted in many post-Soviet countries \cite{ChernenkoWhoTrusts2025, WijermarsTelegramHarbinger2022} and in Iran \cite{GhorbanzadehExaminingTelegram2018, KargarCensorshipCollateral2018} while viewed critically in the West for associations with illegal and extremist content \cite{KiessEuroscepticismLocal2025, WaltherUSExtremism2021}. 

To systematically study discourse on Telegram across these diverse contexts, we collected TeraGram, a large-scale, structured dataset of publicly available Telegram messages using a snowball crawling method~\cite{goodman1961snowball}. This approach systematically discovers and maps interconnected channels and communities, allowing to capture rich temporal and relational data on the platform. 

Several large-scale Telegram datasets have already been published. For example, the Pushshift Telegram dataset \cite{BaumgartnerPushshiftTelegram2020} collected over 300 million messages starting from a seed of extremist and cryptocurrency channels, providing an early glimpse into Telegram’s content landscape. Building on this, TGDataset \cite{LaMorgiaTGDatasetCollecting2025} expanded coverage to 400 million messages across 120,000 channels, offering a more balanced view of the platform without focusing on one topic. Recent efforts have achieved even greater scales. The Telescope dataset features enriched metadata for 500 thousand public channels and message metadata for 71 thousand fully downloaded channels \cite{GangopadhyayTeleScopeLongitudinal2025}. 

In addition to those general-purpose datasets, several datasets focus on specific communities or geopolitical events. A dataset by Blas et al. on the 2024 US Presidential Election \cite{BlasUnearthingBillion2025} includes over one billion messages and uniquely incorporates private groups accessible by invite, with particular emphasis on English-speaking communities. Two datasets examine how the Russia-Ukraine war is covered on Telegram: the dataset by \cite{KireevTelegramDataset2025} features examples of propaganda and moderation, while \cite{BawaTelegramBattlefield2025} contrasts pro- and anti-Kremlin Telegram channels. Finally, the Schwurbelarchiv project \cite{AngermaierSchwurbelarchivGerman2025} collected a smaller number of German channels, and used AI to transcribe multimedia content such as voice messages and videos. A summary of existing datasets is provided in SI, Tab.~\ref{tab:datasets_overview}.

\begin{centeredfigure}
\includegraphics{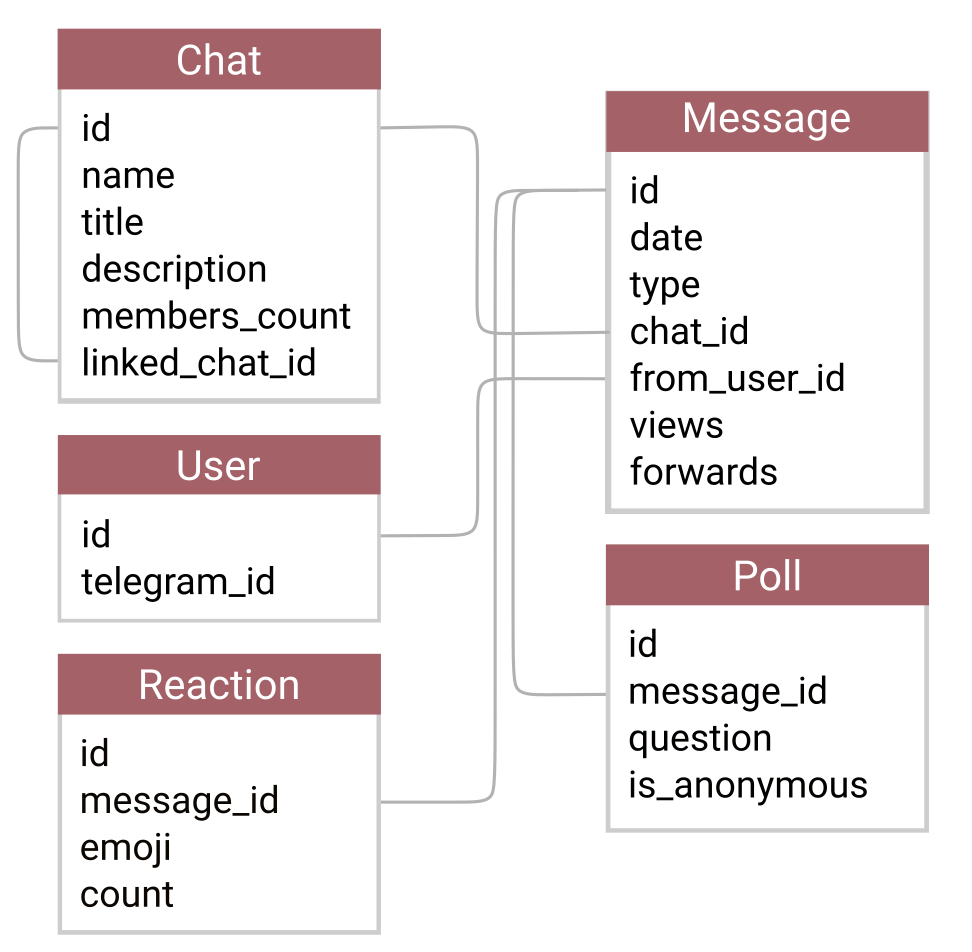}
\captionof{figure}{\captionbf{A simplified entity-relation diagram of the dataset.} See SI, Fig. \ref{fig:er_diagram} for the non-simplified version.}
\label{fig:simplified_er}
\end{centeredfigure}

TeraGram provides the largest longitudinal record of public Telegram content to date (Tab. \ref{tab:summary_statistics}). Beyond its size, it improves over existing datasets in several aspects. First, while most existing datasets are distributed as JSON files that need to be parsed before analysis and might contain inconsistent schemas, TeraGram is structured in a relational format that enables computationally efficient analysis at scale (Fig.~\ref{fig:simplified_er}). The core relational tables Chats, Messages, Users, and Polls are populated with data directly available from the Telegram API, including message metadata (timestamp, view count, forward count), poll questions, and public chat information. Second, TeraGram aims to comprehensively capture Telegram-specific features that are absent or incomplete in prior work, including discussion groups, emoji reactions, polls, embedded URLs, hashtags, and reply-thread relationships. While some of these are available in prior work, no existing dataset provides all of them together. Third, rather than targeting a specific language or community, TeraGram spans a wide range of languages and public communities; we further augment the data by algorithmically inferring the primary language of each chat, enabling systematic cross-lingual analysis.

To preserve privacy, all user identifiers are pseudonymized, only public channel and group metadata is retained, and binary media blobs are excluded. The full message text is available to qualified researchers upon reasonable request.

With a total volume of 3.33~TB of data (1.43~TB excluding text) (Tab.~\ref{tab:summary_statistics} and SI, Tab~\ref{tab:message_types}), the dataset is distributed in a Parquet format that can be easily ingested into a relational database\footnote{Full dataset: \url{https://doi.org/10.25625/GDCXQK}}, accompanied by a representative CSV sample for accessibility\footnote{CSV sample: \url{https://zenodo.org/records/18262126}}. Both datasets are shared under the Open Data Commons Attribution (ODC-By) license that allows users to freely share, modify, and use the dataset as long as they attribute it. In this paper, we illustrate the utility of the dataset with preliminary analyses of language distribution, network degree, and topic clustering.

Overall, TeraGram provides a valuable resource for advancing research in social media analysis, online user behavior and computational social science — particularly within a platform characterized by minimal algorithmic interference.

\section{Results}

\subsection{Dataset overview}

Telegram supports two formats for public communication: channels and groups. \textit{Channels} function as broadcasting services where only administrators can publish content, while subscribers receive and read these messages. \textit{Groups}, in contrast, allow all members to post and participate in conversations. Furthermore, a channel can be linked to a dedicated group for its subscribers where members can discuss the channel's content and comment on individual posts. Such a group is termed a \textit{discussion group}. Throughout this paper, we use the terminology from the Telegram API where the term \textit{chats} serves as an umbrella term for both channels and groups.  

\begin{centeredfigure}
        \footnotesize
    \begin{tabularx}{\linewidth}{@{}Xrr@{}}
        \toprule
        \multicolumn{1}{l}{\textup{Table}} & \multicolumn{1}{r}{\textup{Count}} & \multicolumn{1}{r}{\textup{Size}} \\
        \midrule
        Messages        & 5.95B       & \SI{1.0}{\tera\byte}    \\
        Message text (available on legitimate request)  & 5.51B & \SI{1.9}{\tera\byte} \\
                Hashtags        & 498M        & \SI{33}{\giga\byte}     \\
        URLs            & 655M        & \SI{63}{\giga\byte}     \\
        Reactions       & 3.60B       & \SI{319}{\giga\byte}    \\
        Discovered chats (umbrella term for channels and groups)  & 4.54M  & \SI{1.1}{\giga\byte} \\
        Downloaded chats (includes full message history)          & 712k   & \SI{172}{\mega\byte} \\
        Poll questions  & 21.29M      & \SI{5.2}{\giga\byte}    \\
        Poll answers    & 79.44M      & \SI{8.8}{\giga\byte}    \\
        Users (pseudonymized)   & 15.30M  & \SI{10}{\giga\byte} \\
        Channel members         & 3.57M   & \SI{446}{\mega\byte} \\
        \midrule
        \textup{Total}  &      & \textup{\SI{3.33}{\tera\byte}}    \\
        \bottomrule
    \end{tabularx}
    \captionof{table}{\captionbf{Summary of collected Telegram data.} The values correspond to the approximate number of rows and disk size of the corresponding SQL table, including indexing.
    }
    \label{tab:summary_statistics}
\end{centeredfigure}

\begin{figure*}
    \centering
    \includegraphics{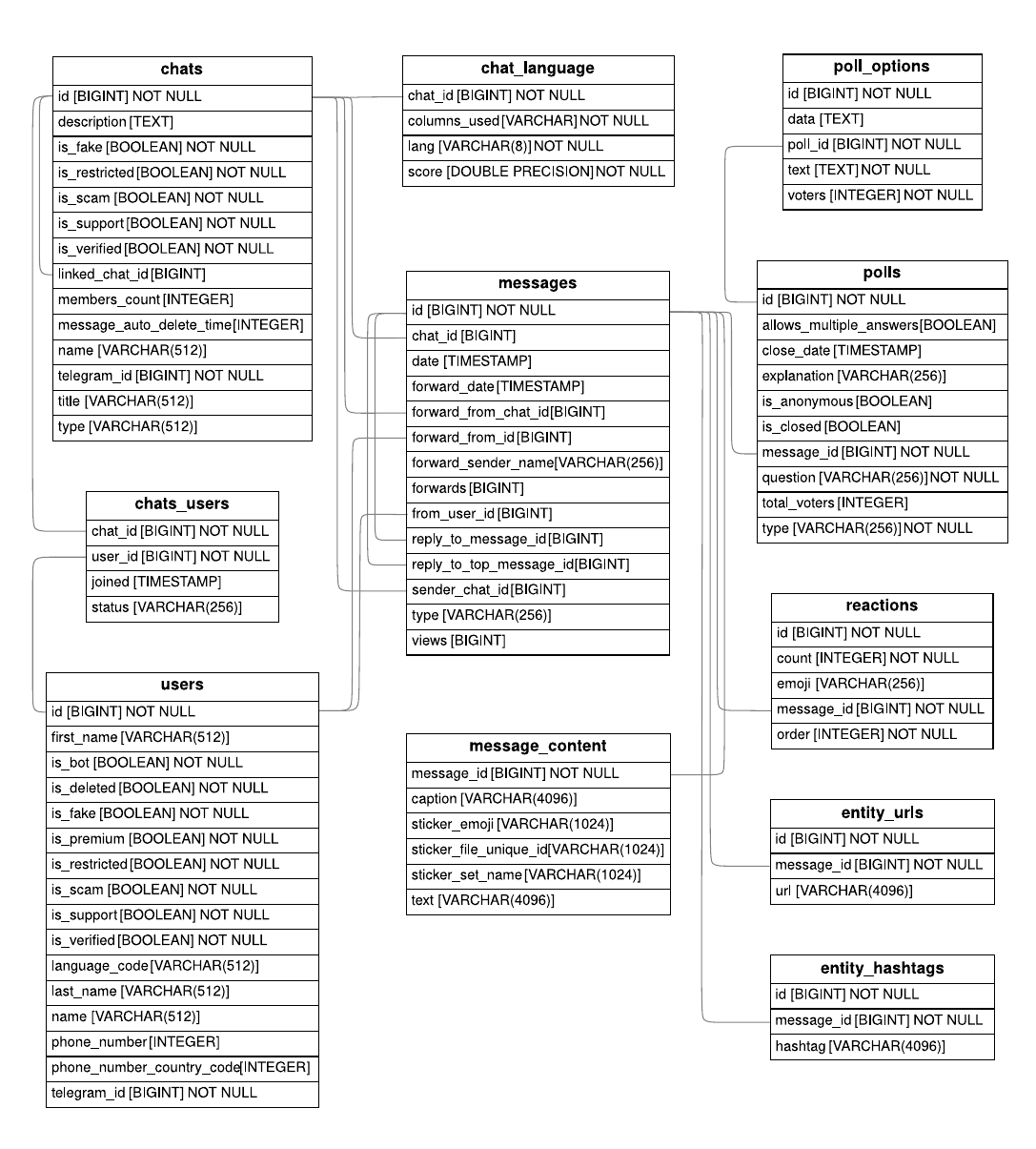}
    \caption{\captionbf{Entity-relation diagram of the SQL database.}}
    \label{fig:er_diagram}
\end{figure*}

Amounting to 3.33 terabyte of data, the TeraGram dataset presented in this study is one of the largest and most comprehensive Telegram datasets available to date (see Tab.~\ref{tab:summary_statistics} for summary statistics). The core dataset is organized into interrelated tables representing key Telegram entities, including messages, users, chats, polls, and emoji reactions (see Fig.~\ref{fig:simplified_er} and SI, Fig.~\ref{fig:er_diagram}). This relational structure enables a detailed reconstruction and analysis of user interactions, group dynamics, and content dissemination patterns within the Telegram ecosystem.

Data was collected via the official Telegram user API using a snowball-crawling approach (see Methods for a detailed description). Starting from the 100 largest political channels by subscriber count (SI, Tab.~\ref{tab:seed_list}), the crawler iteratively discovered new publicly accessible chats through forwarded messages. Discovered chats were prioritized for download based on their out-degree, i.e., the number of times they were forwarded in already downloaded chats. More forwards resulted in higher download priority. This approach ensures that the crawler prioritizes the highly influential chats from the tail of the out-degree distribution, which may present the hubs of the Telegram network (Fig.~\ref{fig:degree_ccdf}). Additionally, focusing on popular chats helps to protect user privacy, as users participating in smaller chats might not expect that their messages, though public, would be visible to a large audience. 

\begin{centeredfigure}
\includegraphics{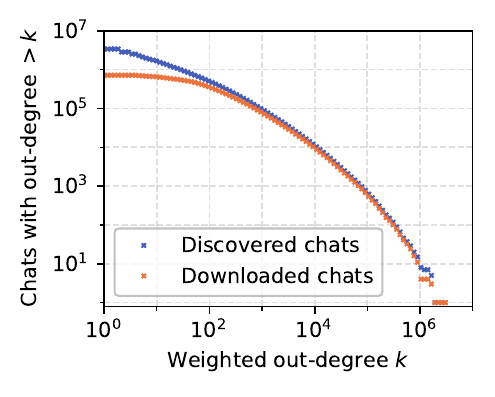}
\captionof{figure}{\captionbf{Complementary cumulative distribution function (CCDF)} of the weighted out-degree of discovered and downloaded chats. The gap between the curves at low degrees is due to the crawler prioritizing chats for download proportionally to the degree. The out-degree is defined as the number of messages forwarded from a given chat to other downloaded chats, with edge weights proportional to the number of forwarded messages.}
\label{fig:degree_ccdf}
\end{centeredfigure}

Overall, we discovered \num{4382659} channels and \num{160617} groups, and fully downloaded \num{711782} channels and \num{10060} groups. Out of those, 4017 groups (40\%) are discussion groups linked to channels. We also collected user profile data from users who posted in groups and from groups where the subscriber list was publicly visible. In those cases, we hashed usernames and first and last names prior to storage to ensure that this sensitive information is not retained at any point during the data collection process. When a user's phone number was visible, we retained only the country code and removed all other digits.

The data was collected between May 2025 and November 2025; however, since the Telegram API provides the whole history of a chat, the collected data spans from September 2015 to November 2025 --- over a decade-long period (Fig.~\ref{fig:messages_per_week}). Because chats were not updated after their initial download, chats encountered later in the crawling process contain more recent messages than those downloaded at the beginning of the collection period. 

\begin{centeredfigure}
\includegraphics{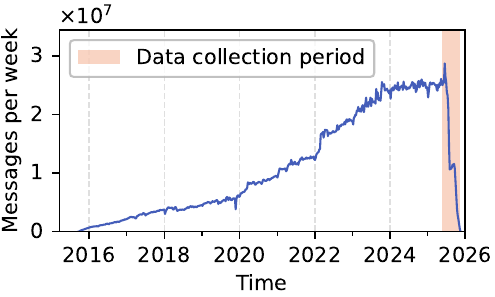}
\captionof{figure}{\captionbf{Number of messages posted per week in the dataset.} The time series spans September 2015 to November 2025. Apparent variations during the collection period are influenced by the crawling procedure: chats were downloaded only once, so those discovered later in the crawl contribute more recent messages than chats collected earlier.}
\label{fig:messages_per_week}
\end{centeredfigure}

The dataset is shared under a two-tiered access model: rich metadata is openly available, while message text is accessible to researchers upon reasonable request. The metadata includes view counts and forward counts, which serve as measures of user engagement. When a message is forwarded from another chat, we link it to its original source message and originating channel when this information is available. Similarly, replies include references to the messages they respond to, creating explicit conversational threads. In addition, we also extracted two types of text markup entities: URLs and hashtags. These provide an efficient way to access structural text information without parsing the raw text. For messages containing binary data such as images, audio messages, and videos, we store only the metadata and any accompanying text captions. 

We also collected data on special features of Telegram such as polls and emoji reactions. Poll data include the question text, answer options, and aggregate vote counts when available. Reactions are stored as aggregated emoji counts per message without user-level information. Message reactions can be used as a measure of user engagement or as an efficient way to evaluate sentiment of a given message. 

\subsection{Quality control}
\label{sec:qc}

Given the scale and diversity of our Telegram corpus, rigorous quality control is essential to ensure data integrity. As outlined below, our quality control workflow combines automated integrity inspection with manual spot-checks on a stratified random sample of messages. We also publish a datasheet in accordance with~\cite{GebruDatasheetsDatasets2021} that provides comprehensive documentation of dataset provenance. Together, those steps verify data integrity and quality before any downstream analysis. 

\subsubsection{Data integrity}

Following the approach recommended by~\cite{ElazarWhatsMy2024}, we execute three automated quality control checks: duplicate detection, length distribution analysis, and n-gram inspection. The goal of those checks is to catch scraping artifacts, duplicates indicating spam or bot activity, and anomalies in message length distributions that can be due to automated truncation. 

Duplicate detection analysis reveals that 53.7\% of messages are unique. Duplications can be explained by several factors: very short messages (e.g., single-word responses or greetings) that naturally occur frequently; messages containing only URLs without additional text; or advertisement posts that are frequently reposted without changes in content. Duplicates were not removed from the dataset.

Length distribution analysis shows several strong peaks at specific message lengths (Fig. \ref{fig:message_length_dist}). We investigated the origin of the four most prominent peaks. The peak at 28 characters (orange star) corresponds to YouTube links that follow the format \enquote{https://youtu.be/abcdefghijk}. The peak at 80 characters (purple square) is due to messages redacted by Telegram for copyright or Terms of Service violations. In such cases, the original message text is replaced by either \enquote{This message couldn't be displayed on your device due to copyright infringement.} or \enquote{This channel can't be displayed because it violated Telegram's Terms of Service.} Both are 80 characters long. The peak at 288 characters (yellow diamond) was caused by an advertisement for other Telegram channels that was mass-posted in one Arabic channel. Finally, the peak at 1024 characters (red triangle) corresponds to the caption length limit, i.e., the message length limit for messages containing multimedia content.

\begin{centeredfigure}
\centering
\includegraphics{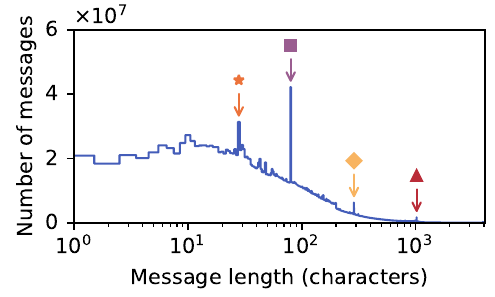}
\captionof{figure}{\captionbf{Distribution of the message length shows several distinct peaks caused by systematic patterns.}  \captionbf{Star}: YouTube links; \captionbf{square}: message redacted by Telegram for copyright or Terms of Service violation; \captionbf{diamond}: an advertisement message mass-posted in one Arabic channel; \captionbf{triangle}: caption length limit.}
\label{fig:message_length_dist}
\end{centeredfigure}

Finally, we performed n-gram analysis to detect artifacts in text such as unusual punctuation, spam, and near-duplicate messages. For this, we extract the top 10 most popular unigrams, 3-grams, and 10-grams from all English-speaking chats (SI, Tables~\ref{tab:unigrams}-\ref{tab:10-grams}). Following the approach in \cite{ElazarWhatsMy2024}, we did not filter stop words or clean the text beyond converting it to lowercase, as the goal is to detect artifacts rather than describe message content. The most popular n-grams show a mix of typical English phrases, especially in unigrams and 3-grams, and inorganic strings that become more common in 10-grams. Those inorganic strings are due to boilerplate text that some channels append to every message, text separators, or redaction messages from Telegram discussed earlier. We also find an additional redaction message: \enquote{This channel can't be displayed because it violated local laws.} This text is displayed whenever a message was forwarded from a chat that violated local laws to a normal chat. Such chats are called restricted in the Telegram API and are marked in the dataset as such in the field \verb|is_restricted|. 

\subsubsection{Bot activity}
A key quality concern is the prevalence of bot-generated messages, as automated posts can skew analyses of human behavior and information spread. Registered Telegram bots, easily identifiable via the API, constitute just \SI{0.4}{\percent} of users and contribute \SI{0.14}{\percent} of messages in our dataset. More insidious \enquote{troll farm} or LLM-driven accounts that mimic human activity are harder to detect and beyond this paper's scope (see, e.g., studies on coordinated inauthentic behavior \cite{CinelliCoordinatedInauthentic2022}). Nevertheless, we do not expect such accounts to be ubiquitous in the dataset. First, our data span back to 2015, predating the recent LLM boom. Second, the dataset primarily consists of channels where posting is restricted to administrators. Consequently, the most significant risk of bot contamination stems from channel administrators using bots to manage their channel. We leave the investigation of this scenario for dedicated future work. 

\subsection{Analysis of the Telegram dataset}

\begin{figure*}[t]
    \includegraphics{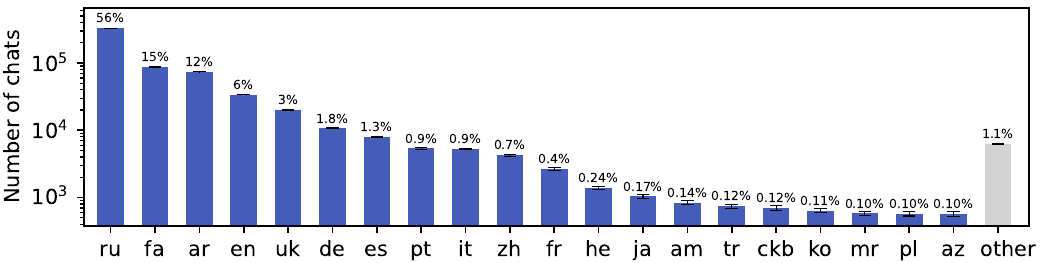}
    \caption{\captionbf{Languages of fully downloaded chats.} Chat language is classified based on the text of the first 100 messages. Language codes follow the ISO standard. Error bars give the 95\% CI interval.}
    \label{fig:language_distribution}
\end{figure*}

To characterize the thematic and informational range of the dataset, we conduct analyses focusing on language use, external references, and topical structure.

We identify the primary language of each downloaded chat using the \verb|fast_langdetect| library \cite{LlmKiraFastlangdetect, JoulinBagTricks2016, JoulinFastTextzipCompressing2016}. For each chat, we use the first 100 messages since creation, all concatenated into one long string, where we remove URLs and whitespace. Chats with a language-classification confidence score below 0.8 are excluded from the analysis. Error bars on language frequency counts are computed using Wilson score confidence intervals, treating each language count as a binomial proportion of the total corpus.

The dataset covers a wide range of languages, with top-5 languages accounting for approx. \SI{91.7}{\percent} of all chats (Fig.~\ref{fig:language_distribution}). We find a high fraction of Russian (\SI{56}{\percent}) and Farsi (\SI{15}{\percent}) chats, which reflects Telegram’s mainstream adoption in Russia and other post-Soviet countries and Iran. English chats make \SI{6}{\percent} of downloaded chats, which still amounts to \num{33829} chats. 

To evaluate the quality of our language classification, we analyzed whether forwarded messages typically originate and arrive within a chat of the same language (e.g., Russian to Russian). This is the case for 83.3\% of forwards where the source chat was successfully downloaded. In almost half the cases where source and destination were classified differently, either the source (25\%) or destination (19\%) was English. The former fraction likely represents non-English chats forwarding English material, while the latter, perhaps, represents chats that write in English even though they belong to a non-English subnetwork. As a further control, we manually inspected sets of 100 random messages from Russian, English, Arabic and German chats and found a low number of misclassifications (5\%, 9\%, 1\% and 3\%, respectively), consistent with the reported 95\% accuracy of \verb|fast-langdetect| \cite{LlmKiraFastlangdetect, JoulinBagTricks2016, JoulinFastTextzipCompressing2016}.

To evaluate the prevalence of misinformation in the data, we assessed the reliability of news domains shared in messages using the aggregated \enquote{wisdom of experts} domain rating by~\cite{LinHighLevel2023}. Since the domain rating primarily covers the US media landscape and has limited coverage of Russian and Iranian news sources, we only performed this analysis on English-language chats. For every chat, we extracted all URL entities from the corresponding table. URL rehydration was not necessary, as link shorteners are rarely used on Telegram; for example, Bitly links make up only 1.2\% of URLs. We applied a blacklist to remove domains like search engines and social media platforms that appear in the Lin et al.\ dataset but do not refer to news sources (SI, Tab.~\ref{tab:blacklist}). Of the remaining URLs, we randomly sampled 1\% to reduce computational costs, which resulted in 861~thousand URLs, and fit a kernel density estimator with a Gaussian kernel to estimate the probability distribution. To compute the CI intervals, we perform clustered bootstrapping with domains as clusters; in other words, instead of resampling individual URLs, we sample entire domains with replacement 1000 times. 

\begin{centeredfigure}
    \includegraphics{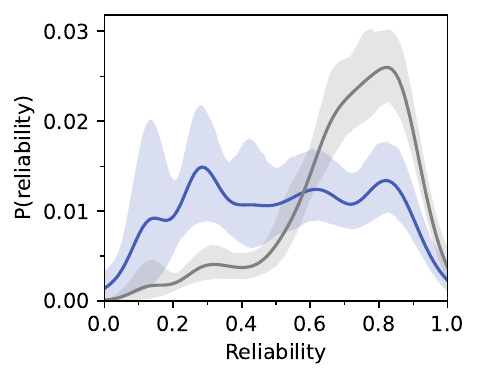}
    \captionof{figure}{\captionbf{High prevalence of unreliable URLs in English-speaking Telegram chats compared to a mainstream platform like Twitter.} The bands represent the 95\% CI interval obtained by clustered bootstrapping on domains.}
    \label{fig:url_reliability}
\end{centeredfigure}

Overall, we observe a high prevalence of URLs with a reliability score below 0.6 in English-speaking Telegram chats (Fig.~\ref{fig:url_reliability}). For comparison, we perform the same analysis on a Twitter dataset of all tweets within a 24-hour period on September 21, 2022 \cite{PfefferJustAnother2023} (Fig.~\ref{fig:url_reliability}, gray line). The resulting distributions differ distinctly: URLs on Twitter are concentrated within the 0.6 to 1.0 reliability range. In contrast, the distribution for English-speaking Telegram varies widely, with 60\% of URLs lying below the 0.6 threshold. 

To explore the main themes discussed in chats, we applied BERTopic \cite{GrootendorstBERTopicNeural2022}, an unsupervised topic modeling pipeline, to messages from the four most prevalent languages in our dataset: Russian, Farsi, Arabic, and English. For each language, we filtered chats with a language classification score above 0.8 and extracted all messages from those chats. We cleaned the messages by removing URLs and stripping consecutive whitespace, then filtered out messages shorter than 50 characters and system messages identified in quality control analysis. We then randomly sampled 1 million messages per language and performed topic modeling on each set, with each run being executed on an NVIDIA A100 40GB GPU. To produce readable labels for each topic, we use BERTopic's ChatGPT integration (\verb|gpt-4o-mini| model), where we provide it with topic keywords identified by BERTopic and a sample of messages and ask it to return a short label in English. The preprocessing code and topic modeling pipeline with all hyperparameters are available in our GitHub repository.

%\openwidetext
    \begin{figure*}[t]
    \includegraphics{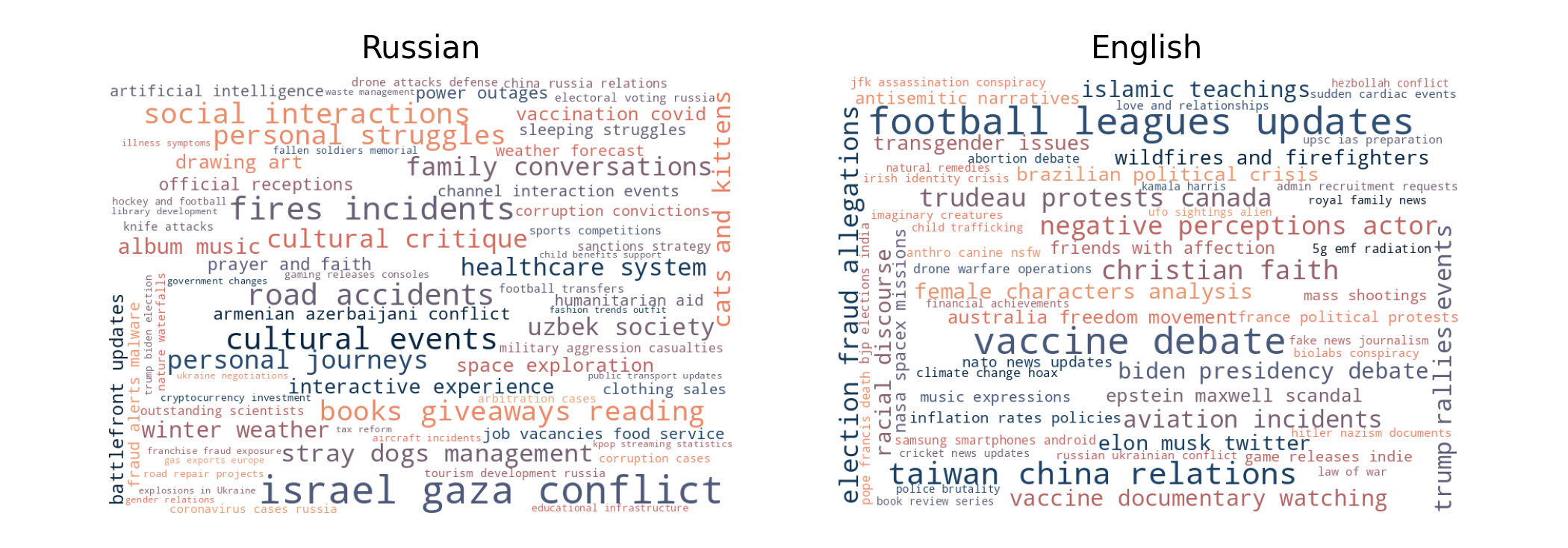}
    \captionof{figure}{\captionbf{Topics in Russian and English chats identified using BERTopic.} While both English and Russian datasets contain topics like sports and current events, the English dataset includes a subset of far-right topics (e.g., \enquote{antisemitic narratives,} \enquote{climate change hoax}). In contrast, Russian topics predominantly reflect mainstream diverse interests, including books, fashion, art, and music. See Tables~\ref{tab:topics_ru}-\ref{tab:topics_en} in the SI for the top-40 topics of the four languages in a tabular format.}
    \label{fig:topics_ru_en}
    \end{figure*}
% \closewidetext

\begin{figure*}
    \centering
    \includegraphics[width=0.9\linewidth]{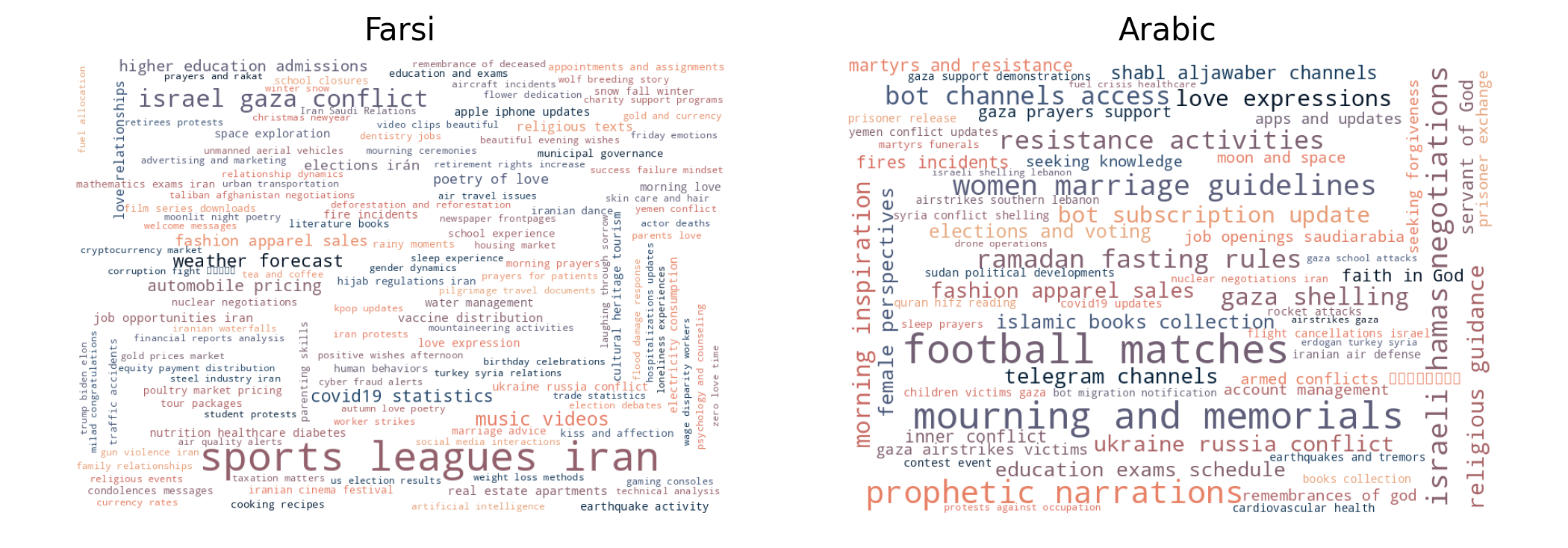}
    \caption{\captionbf{Topics in Farsi and Arabic chats identified using BERTopic.}}
    \label{fig:topics_fa_ar}
\end{figure*}

All four languages feature discussions about politics, current world events, as well as non-political topics such as sports and music (Fig.~\ref{fig:topics_ru_en}, SI, Fig.~\ref{fig:topics_fa_ar}, and SI Tables~\ref{tab:topics_ru}-\ref{tab:topics_en} for the top-40 topics in tabular format).
However, English top-40 topics also include entries like \enquote{racial discourse}, \enquote{antisemitic narratives}, \enquote{climate change hoax} and \enquote{jfk assassination conspiracy} that suggest far-right and conspiracy content. In contrast, Russian and Farsi topics cover a more diverse set of day-to-day topics like books, fashion, art, and music, whereas Arabic topics are predominantly religious.

\subsection{Usage guide}
The dataset is designed for flexible use across many fields, including computational social science, digital humanities, and machine learning applications. It is provided as a relational database, with tables representing messages, channels, users, polls, and reactions. This structure supports efficient queries on content, metadata, and user-channel interactions.

The dataset adheres to FAIR principles~\cite{wilkinson2016fair} to ensure researchers can seamlessly download and integrate the data into existing workflows:
\begin{itemize}
\item \textbf{Findable:} Both the full dataset and subsampled CSV tables are indexed with DOIs on established data-sharing platforms.
\item \textbf{Accessible:} Public portions of the dataset are available for download without authorization.
\item \textbf{Interoperable:} The full dataset is provided as Parquet files, which is an open-source table format supported by many data processing systems. The subsampled dataset is provided in CSV format for readability and easy inspection.
\item \textbf{Reusable:} To facilitate reuse, we provide schema documentation, access instructions, example SQL queries, and preprocessing scripts in our GitHub repository\footnote{\url{https://github.com/Priesemann-Group/telegram_quality_control}}. Dataset provenance and limitations are documented in an accompanying datasheet.
\end{itemize}

Ethical use is strongly encouraged. Researchers are advised to respect the original context of communication and avoid deanonymization efforts. 
    
\section{Discussion}

In this work, we present TeraGram, a large-scale structured dataset of the publicly available Telegram ecosystem. Our dataset is, to the best of our knowledge, the largest to date; it also contains rich metadata on various features of Telegram such as polls, emoji reactions, URLs, hashtags, discussion groups, forwards and replies. Message texts are available upon request. The structured format provides a convenient and efficient way for researchers to analyze various aspects of the platform. 

The dataset spans a wide range of languages, with billions of messages in Russian, Farsi, and Arabic, and hundreds of million in English and other European languages. The high proportion of Russian and Farsi content reflects Telegram's popularity in Russia and other post-Soviet countries and Iran. 

Telegram usage varies across languages. While Russian and Farsi chats feature more mainstream topics, English chats share URLs of remarkably low reliability: 59\% fall below the commonly accepted 0.6 threshold for classification as reliable sources. This contrasts with mainstream English social media, where a comparable study found only about 15\% of sources to be questionable \cite{DiMartinoIdeologicalFragmentation2025}. Direct comparison for Russian and Farsi is not currently possible, as existing reliability rankings lack sufficient coverage of Russian and Iranian news sources.

Several important limitations must be considered when working with this data. First, since the crawler prioritizes queued chats based on the number of messages that were forwarded from them, it is biased towards popular chats. We also cannot reach chats that do not belong to the same connected component as our seed chats. Additionally, we did not crawl private conversations and groups. Researchers should therefore exercise caution when extrapolating findings to broader user behavior on Telegram, especially in contexts where private or low‑visibility conversations play a critical role. 

Second, Telegram itself has undergone significant evolution during the nearly decade-long period that our data covers. Early on, the platform operated with minimal algorithmic intervention, but in 2024, features such as recommendations of related channels, sponsored messages, and enhanced content‑discovery tools have been introduced \cite{telegramChangelog}. This evolution implies that interaction dynamics likely differ between 2017 and 2025. Moreover, natural shifts in language and community norms over time introduce additional variance. Consequently, any analysis pooling data across this period must account for this temporal heterogeneity.

Overall, the presented dataset allows a wide range of future investigations into online communication and information dynamics. For instance, recent work  already leveraged the data to study the spread of information via external URLs \cite{Ventzke2025}, fine-tune LLMs \cite{BrockersDisentanglingInteraction2025}, or improve misinformation detection algorithms \cite{KesslerNetwork}.

We anticipate that the dataset will support diverse downstream applications, including network modeling, bot detection, and community formation across multilingual and longitudinal corpora.

\section{Conclusion}

In this paper, we collected a longitudinal dataset of public Telegram chats spanning nearly a decade worth of data. The dataset thus provides a unique opportunity to study organic information diffusion across diverse community structures on a platform with minimal moderation and algorithmic interference.

\section{Methods}

\subsection{Data collection}

Telegram provides no official data access for researchers, necessitating a custom collection pipeline. We built an asynchronous crawler using Pyrogram~\cite{pyrogram}, a Python interface to the MTProto Telegram API, which interacts with the platform through standard client authentication.

We employed a snowball crawling strategy~\cite{goodman1961snowball} that exploits Telegram's message-forwarding feature to discover interconnected chats. Starting from a curated seed set of 100 public channels (see SI, Tab.~\ref{tab:seed_list}), the crawler recursively identified new chats by resolving the origins of forwarded messages. This approach enables scalable, structurally informed data collection~\cite{LaMorgiaTGDatasetCollecting2025,BaumgartnerPushshiftTelegram2020}.

To maximize coverage of influential chats, we assigned a download priority to queued chats equal to their observed out-degree, i.e., the number of times the crawler encountered forwards from the chat in already downloaded chats. The crawler always downloaded the chat with the highest priority of all discovered and not yet downloaded chats, ensuring central network hubs were captured first (SI, Algorithm~\ref{alg:snowball_crawling}). This priority queue approach optimizes the discovery of large, interconnected chats within sampling constraints.

The crawler implemented several specific behaviors to ensure data completeness. For each channel, we also crawled its linked discussion group (if it existed), capturing user comments and interactions that provide crucial context for message interpretation. To access statistics of non-closed polls, we had to cast a vote before recording the results, then subtracted our own vote during post-processing to maintain the original voting distributions. Additionally, we addressed two data integrity challenges. First, when reconstructing reply threads, we could not rely on Telegram's sequential message identifiers, because those shift when messages are deleted to preserve consecutive indexing. Instead, we matched replies using timestamps. Second, we encountered the same problem when identifying the original source of forwarded messages, where we also matched the messages by timestamps. This approach might have led to inconsistencies whenever two messages are posted in the origin chat within the same second. 

Telegram imposes rate limits on API requests per account. To scale data collection, we distributed the crawling process across 200 authenticated accounts managed by three worker machines. This parallelization enabled us to bypass per-account limitations while maintaining a single database instance for consolidated storage, running on a fourth separate machine with local NVMe SSD storage drives.

We collected comprehensive metadata including messages, channels, users, polls, and forwarding relationships (SI, Tab. \ref{tab:message_types}). Binary content (images, videos) was excluded due to storage and copyright considerations. Phone numbers were excluded due to privacy concerns. Usernames were hashed upon storage. The resulting dataset was stored in a PostgreSQL relational database optimized for high-volume insert performance.

\section*{Acknowledgments} 

We would like to thank Jana Lasser for providing early access to the Twitter dataset used in this study.
Authors with affiliation \enquote{1} received support from the Max-Planck Society. A.G., S.B.M., A.I.G., and V.P. were funded by the German Federal Ministry for Education and Research for the infoXpand project (031L0300A). A.I.G. and A.C.S. were funded by the German Research Foundation – GRK2906 – project number 502807174. U.H. was funded by the Danish National Research Foundation (grant no. DNRF170). J.P.N. was funded by the Austrian Science Fund (DOI:10.55776/P37280). V.P. and A.C.S. were supported by the German Research Foundation under Germany’s Excellence Strategy–EXC 2067/1-390729940 (MBExC). This work was further supported by the MWK Niedersachsen via the programs \enquote{zukunft.niedersachsen}, \enquote{Niedersächsisches Vorab} and \enquote{Niedersachsen-Profil-Professur}.

ChatGPT, Claude, and Qwen3 were used for proofreading and minor stylistic corrections. In the topic modeling analysis, we used BERTopic's ChatGPT integration to generate readable English labels for the identified topics. 

\printbibliography

\end{multicols}

\newpage

\pagestyle{plain}

\restoregeometry
\newgeometry{
    left=1.55cm,
    right=1.55cm,
    top=1.5cm,
    bottom=2cm,
}

\section*{Appendix}

\FloatBarrier

\renewcommand{\arraystretch}{1.5}

\begin{table*}[h]
    \centering
    \footnotesize
    \begin{tabularx}{\textwidth}{p{0.1\textwidth} p{0.13\textwidth} p{0.10\textwidth} p{0.15\textwidth} p{0.08\textwidth} p{0.12\textwidth} Y@{}}
        \toprule
        \textbf{Dataset} & \textbf{Size} & \textbf{Timespan} & \textbf{Topic focus} & \textbf{Format} & \textbf{Includes text} & \textbf{Features} \\
        \midrule
        TeraGram & 712k chats, \newline 5.95B messages & Sep. 2015 – \newline Nov 2025 & General purpose & Parquet & On request & Discussion groups, reply trees, polls, emoji reactions \\
        Blas et. al & 43k chats, \newline 1B messages & Aug. 2024 – \newline Feb. 2025 & US elections & SQLite & Yes & Link-accessible private chats \\
        TGDataset & 120k channels, \newline 400M messages & Jan. 2021 – \newline Jul. 2022 & General purpose & JSON & Yes & \\
        Pushshift Dataset & 27.8K chats, \newline 317M messages & Sep. 2025 – \newline Nov. 2019 & Seed: right-wing \newline extremism, crypto & JSON & Yes & \\
        TeleScope & 71k chats, \newline 120M messages & 2015 – \newline Oct. 2024 & General purpose & JSON, \newline CSV & No & Temporal message posting patterns, extracted entities \\
        Schwurbelarchiv & 6k chats, \newline 40M messages, \newline 3M audio files & Oct. 2015 – \newline Jul. 2022 & German-language conspiracies & CSV & Yes & Transcribed multimedia content \\
        Kireev et. al & 13 channels,\newline 17.3M messages & Oct. 2020 – \newline Jan. 2024 & Russian-Ukrainian propaganda & CSV & Yes &  Real-time data collection \\
        Bawa et. al & 519 channels, \newline 5.2M Messages & Dec. 2020 – \newline Apr. 2023 & Pro- vs. anti-Kremlin stances & CSV & Yes & Replies, emoji reactions \\
        \bottomrule
    \end{tabularx}
    \caption{\captionbf{A comparison of TeraGram and existing Telegram datasets.} The number of chats refers to the number of fully downloaded chats.}
    \label{tab:datasets_overview}
\end{table*}

\FloatBarrier

\renewcommand{\arraystretch}{1.2}

\begin{table*}
    \centering
    \footnotesize
    \begin{tabularx}{\textwidth}{@{}l r l Y@{}}
        \toprule
        \textbf{Type} & \textbf{Fraction (\%)} & \textbf{Content Included} & \textbf{Reason for Omission (if any)} \\
        \midrule
        TEXT         & 30.95      & Yes, upon legitimate request         & - \\
        PHOTO        & 27.97      & No  (caption on request)             & Binary media excluded; only textual metadata retained \\
        VIDEO        &  7.92      & No  (caption on request)             & Binary media excluded \\
        WEB\_PAGE    &  5.07      & Yes (link text and metadata)         & - \\
        AUDIO        &  1.62      & No  (caption on request)             & Binary media excluded \\
        STICKER      &  1.55      & No  (identifier only)                & Media omitted; only human-readable emoji label retained \\
        DOCUMENT     &  1.27      & No  (caption on request)             & Binary media excluded \\
        ANIMATION    &  1.17      & No                                   & Binary media excluded \\
        VOICE        &  0.44      & No  (caption on request)                   & Binary media excluded \\
        VIDEO\_NOTE  &  0.41      & No  (caption on request)                   & Binary media excluded \\
        POLL         &  0.25      & Yes (question and options)           & Vote counts may be partially hidden \\
        DICE         & $<$0.01    & No                                   & Content depends on server-side evaluation; not reproducible \\
        LOCATION     & $<$0.01    & No                                   & Location data excluded due to privacy concerns \\
        CONTACT      & $<$0.01    & No                                   & Contact information excluded due to privacy concerns \\
        VENUE        & $<$0.01    & No                                   & Venue/location data excluded due to privacy concerns \\
        GAME         & $<$0.01    & No                                   & Server-side logic unsupported; content not retrievable \\
        PAID\_MEDIA  & $<$0.01    & No                                   & Binary/media excluded; restricted content \\
        GIVEAWAY     & $<$0.01    & No                                   & Binary/media or server-side logic excluded \\
        \bottomrule
    \end{tabularx}
    \caption{\captionbf{Message types in the dataset.} For each type, we show the total count, share of the dataset, what information was retained, and a brief justification for any omission. Binary media and interactive or privacy-sensitive content were excluded to ensure compliance and scalability.}
    \label{tab:message_types}
\end{table*}

\begin{figure*}
    \centering
    \includegraphics{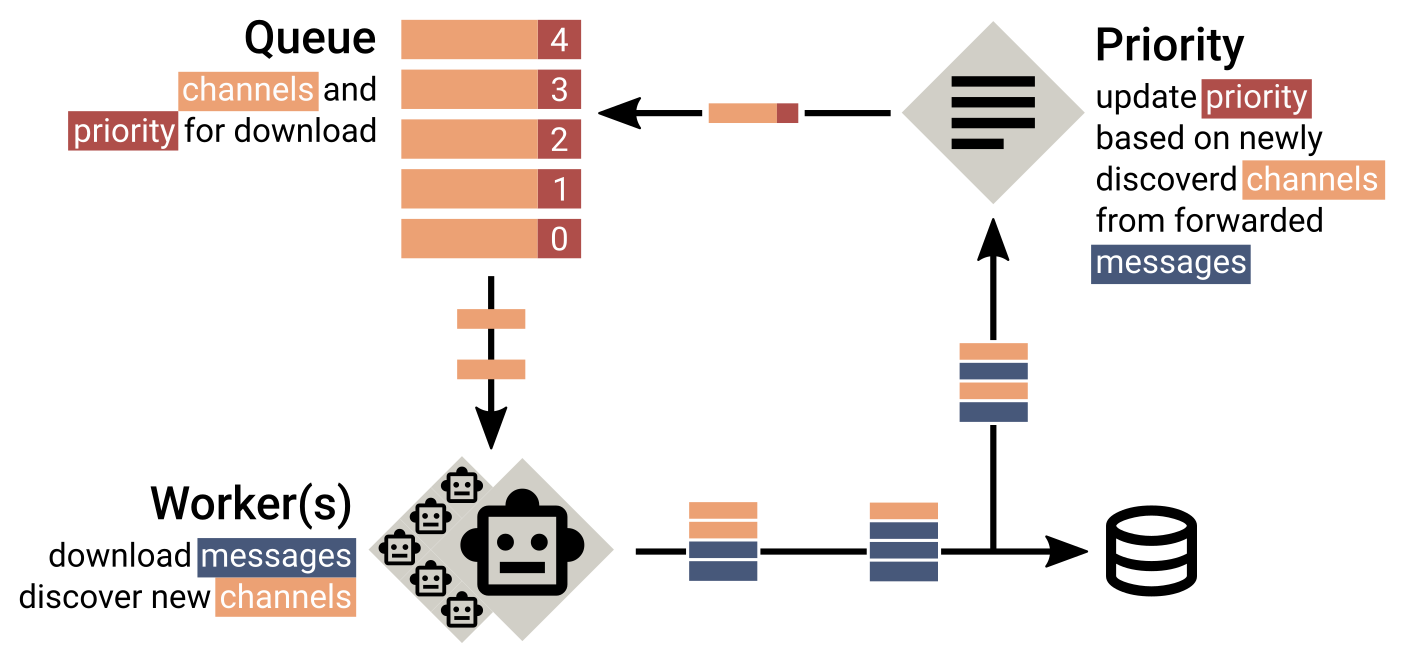}
    \caption{\captionbf{A sketch of the crawling algorithm.} The crawler discovers new chats through forwarded messages. The chats are then prioritized for download based on their out-degree, i.e., the number of forwarded messages from this chat into already downloaded chats.}
    \label{fig:guidance}
\end{figure*}

\FloatBarrier

\begin{table*}
    \centering
    \footnotesize
    \begin{tabularx}{\textwidth}{@{}l X}
        \toprule
        \textbf{Unigram} & \textbf{Approximate count} \\
        \midrule
        \verb"the"& 303780793 \\
        \verb"to" & 194300070 \\
        \verb"of" & 139954153 \\
        \verb"and" & 138496907 \\
        \verb"a" & 108742672 \\
        \verb"in" & 106484729 \\
        \verb"on" & 86053545 \\
        \verb"is" & 78172495 \\
        \verb"this" & 77054100 \\
        \verb"be" & 67760360 \\
        \bottomrule
    \end{tabularx}
    \caption{Top-10 unigrams.}
    \label{tab:unigrams}
\end{table*}

\begin{table*}
    \centering
    \footnotesize
    \begin{tabularx}{\textwidth}{@{}l X}
        \toprule
        \textbf{3-gram} & \textbf{Approximate count} \\
        \midrule
            \verb"due to copyright" & 36559249 \\
            \verb"one of the" & 1739521 \\
            \verb"- - -" & 1557765 \\
            \verb"the united states" & 1247866 \\
            \verb"this is the" & 1185023 \\
            \verb"a lot of" & 1051740 \\
            \verb"this is a" & 998812 \\
            \verb"part of the" & 871766 \\
            \verb"be able to" & 846558 \\
            \verb"as well as" & 833266 \\
        \bottomrule
    \end{tabularx}
    \caption{Top-10 3-grams. We removed all 3-grams that are substrings of the Telegram redaction message on copyright violation except the first one.}
    \label{tab:3-grams}
\end{table*}

\begin{table*}
    \centering
    \footnotesize
    \begin{tabularx}{\textwidth}{@{}l X}
        \toprule
        \textbf{10-gram} & \textbf{Approximate count} \\
        \midrule
        \verb"message couldn't be displayed on your device due to copyright" & 36553503 \\
        \verb"- - - - - - - - - -" & 1045726 \\
        \verb"| gettr | truthsocial | rumble | telegram " \emoji{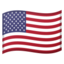} \emoji{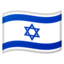} & 302400 \\
        \verb"north atlantic treaty organization by @USERNAME a @USERNAME project -" & 278543 \\
        \verb"can’t be displayed because it violated telegram's terms of service." & 207879 \\
        \verb"subscribe " \emoji{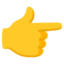} \verb"https://t.me/USERNAME franksocial | gettr | truthsocial | rumble" & 185394 \\
        \verb"help us spread truth! " \emoji{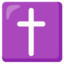} \emoji{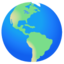} \emoji{us} \verb"main: t.me/USERNAME news: t.me/USERNAME videos:" & 181626 \\
        \verb"support team is here to help with your trb product-related" & 153010 \\
        \verb"this channel can’t be displayed because it violated local laws." & 104086 \\
        \cyrillicdivider  & 89937 \\
        \bottomrule
    \end{tabularx}
    \caption{Top-10 10-grams. Many 10-grams discovered by the analysis were shifted substrings of the same message. In such cases, we show only the first occurrence for readability. Text in capital letters marks redacted content.}
    \label{tab:10-grams}
\end{table*}

\FloatBarrier

\begin{table*}
   \centering
   \footnotesize
    \begin{tabularx}{\textwidth}{@{}l X}
        \toprule
        \textbf{Category} & \textbf{Domain} \\
        \midrule
        Social media & facebook.com, twitter.com, instagram.com, linkedin.com, youtube.com, youtu.be, twitch.tv, reddit.com, redd.it, pinterest.com, tiktok.com, snapchat.com, whatsapp.com, telegram.com, telegram.org, discord.com, viber.com, line.me, kakaotalk.com, wechat.com, qq.com, weibo.com, t.me, t.co, x.com, substack.com, parler.com, rumble.com, gab.com, truthsocial.com, 4chan.org, bitchute.com, gettr.com, dlive.tv, odysee.com, wa.me \\
        \midrule
        Search engines & google.com, bing.com, yahoo.com, duckduckgo.com, baidu.com, yandex.com, goo.gl \\
        \midrule
        E-commerce & amazon.com, amzn.to, amazon.in, paypal.me, paypal.com \\
        \midrule
        Link shorteners & bit.ly, rebrand.ly \\
        \midrule
        Misc. & wordpress.com, lbry.com, minds.com, change.org, gmx.net, soundcloud.com, doi.org, github.com, ift.tt, linktre.ee \\
        \bottomrule
    \end{tabularx}
    \caption{Blacklist of domains excluded from URL classification.}
    \label{tab:blacklist}
\end{table*}

\renewcommand{\arraystretch}{1}

\begin{table*}
    \centering
    \footnotesize
    \begin{tabularx}{\textwidth}{l l l l}
        \toprule
        \textbf{Topic id} & \textbf{Topic} & \textbf{Message count} & \textbf{Fraction} \\
        \midrule
-1	&	political economy russia	&	711168	&	0.711168 \\
0	&	israel gaza conflict	&	8263	&	0.008263 \\
1	&	fires incidents	&	5402	&	0.005402 \\
2	&	cultural events	&	4767	&	0.004767 \\
3	&	road accidents	&	4580	&	0.00458 \\
4	&	social interactions	&	4377	&	0.004377 \\
5	&	books giveaways reading	&	4166	&	0.004166 \\
6	&	personal struggles	&	3946	&	0.003946 \\
7	&	cultural critique	&	3580	&	0.00358 \\
8	&	anonymous messages	&	3580	&	0.00358 \\
9	&	personal journeys	&	3477	&	0.003477 \\
10	&	family conversations	&	3472	&	0.003472 \\
11	&	cats and kittens	&	3319	&	0.003319 \\
12	&	stray dogs management	&	3264	&	0.003264 \\
13	&	healthcare system	&	3125	&	0.003125 \\
14	&	uzbek society	&	3038	&	0.003038 \\
15	&	winter weather	&	2702	&	0.002702 \\
16	&	album music	&	2516	&	0.002516 \\
17	&	interactive experience	&	2359	&	0.002359 \\
18	&	drawing art	&	2247	&	0.002247 \\
19	&	space exploration	&	2242	&	0.002242 \\
20	&	vaccination covid	&	2136	&	0.002136 \\
21	&	battlefront updates	&	2111	&	0.002111 \\
22	&	power outages	&	1817	&	0.001817 \\
23	&	prayer and faith	&	1809	&	0.001809 \\
24	&	official receptions	&	1757	&	0.001757 \\
25	&	armenian azerbaijani conflict	&	1713	&	0.001713 \\
26	&	channel interaction events	&	1625	&	0.001625 \\
27	&	weather forecast	&	1617	&	0.001617 \\
28	&	artificial intelligence	&	1600	&	0.0016 \\
29	&	fraud alerts malware	&	1543	&	0.001543 \\
30	&	job vacancies food service	&	1536	&	0.001536 \\
31	&	sleeping struggles	&	1504	&	0.001504 \\
32	&	humanitarian aid	&	1497	&	0.001497 \\
33	&	corruption convictions	&	1413	&	0.001413 \\
34	&	clothing sales	&	1384	&	0.001384 \\
35	&	military aggression casualties	&	1333	&	0.001333 \\
36	&	outstanding scientists	&	1303	&	0.001303 \\
37	&	sanctions strategy	&	1260	&	0.00126 \\
38	&	china russia relations	&	1253	&	0.001253 \\
        \bottomrule
    \end{tabularx}
    \caption{\captionbf{Top-40 Russian topics}}
    \label{tab:topics_ru}
\end{table*}

\begin{table*}[h]
    \centering
    \footnotesize
    \begin{tabularx}{\textwidth}{l l l l}
        \toprule
        \textbf{Topic id} & \textbf{Topic} & \textbf{Message count} & \textbf{Fraction} \\
        \midrule
-1	&	social issues	&	646264	&	0.646264 \\
0	&	sports leagues iran	&	55357	&	0.055357 \\
1	&	israel gaza conflict	&	12168	&	0.012168 \\
2	&	music videos	&	6586	&	0.006586 \\
3	&	weather forecast	&	5953	&	0.005953 \\
4	&	covid19 statistics	&	5242	&	0.005242 \\
5	&	automobile pricing	&	4650	&	0.00465 \\
6	&	fashion apparel sales	&	4375	&	0.004375 \\
7	&	higher education admissions	&	4163	&	0.004163 \\
8	&	poetry of love	&	3478	&	0.003478 \\
9	&	elections irán	&	3298	&	0.003298 \\
10	&	religious texts	&	3069	&	0.003069 \\
11	&	love relationships	&	3046	&	0.003046 \\
12	&	real estate apartments	&	2707	&	0.002707 \\
13	&	job opportunities iran	&	2527	&	0.002527 \\
14	&	vaccine distribution	&	2485	&	0.002485 \\
15	&	love expression	&	2121	&	0.002121 \\
16	&	water management	&	2119	&	0.002119 \\
17	&	cultural heritage tourism	&	2110	&	0.00211 \\
18	&	morning love	&	2102	&	0.002102 \\
19	&	nutrition healthcare diabetes	&	2020	&	0.00202 \\
20	&	fire incidents	&	1926	&	0.001926 \\
21	&	electricity consumption	&	1844	&	0.001844 \\
22	&	space exploration	&	1787	&	0.001787 \\
23	&	ukraine russia conflict	&	1758	&	0.001758 \\
24	&	apple iphone updates	&	1688	&	0.001688 \\
25	&	earthquake activity	&	1648	&	0.001648 \\
26	&	kiss and affection	&	1496	&	0.001496 \\
27	&	traffic accidents	&	1493	&	0.001493 \\
28	&	poultry market pricing	&	1435	&	0.001435 \\
29	&	morning prayers	&	1397	&	0.001397 \\
30	&	nuclear negotiations	&	1328	&	0.001328 \\
31	&	cooking recipes	&	1299	&	0.001299 \\
32	&	school closures	&	1284	&	0.001284 \\
33	&	iranian cinema festival	&	1276	&	0.001276 \\
34	&	hijab regulations iran	&	1192	&	0.001192 \\
35	&	film series downloads	&	1143	&	0.001143 \\
36	&	condolences messages	&	1140	&	0.00114 \\
37	&	snow fall winter	&	1108	&	0.001108 \\
38	&	rainy moments	&	1097	&	0.001097 \\
        \bottomrule
    \end{tabularx}
    \caption{\captionbf{Top-40 Farsi topics}}
    \label{tab:topics_fa}
\end{table*}

\begin{table*}[h]
    \centering
    \footnotesize
    \begin{tabularx}{\textwidth}{l l l l}
        \toprule
        \textbf{Topic id} & \textbf{Topic} & \textbf{Message count} & \textbf{Fraction} \\
        \midrule
-1	&	religious practices	&	703045	&	0.703045 \\
0	&	football matches	&	13133	&	0.013133 \\
1	&	mourning and memorials	&	10789	&	0.010789 \\
2	&	prophetic narrations	&	7544	&	0.007544 \\
3	&	women marriage guidelines	&	5899	&	0.005899 \\
4	&	israeli hamas negotiations	&	5894	&	0.005894 \\
5	&	resistance activities	&	4820	&	0.00482 \\
6	&	ramadan fasting rules	&	4736	&	0.004736 \\
7	&	bot channels access	&	4353	&	0.004353 \\
8	&	gaza shelling	&	4108	&	0.004108 \\
9	&	morning inspiration	&	3725	&	0.003725 \\
10	&	religious guidance	&	3579	&	0.003579 \\
11	&	bot subscription update	&	3483	&	0.003483 \\
12	&	love expressions	&	3405	&	0.003405 \\
13	&	ukraine russia conflict	&	3157	&	0.003157 \\
15	&	telegram channels	&	3017	&	0.003017 \\
14	&	fashion apparel sales	&	3017	&	0.003017 \\
16	&	education exams schedule	&	2678	&	0.002678 \\
17	&	female perspectives	&	2601	&	0.002601 \\
18	&	islamic books collection	&	2458	&	0.002458 \\
19	&	elections and voting	&	2345	&	0.002345 \\
20	&	servant of God	&	2285	&	0.002285 \\
21	&	shabl aljawaber channels	&	2254	&	0.002254 \\
22	&	inner conflict	&	1903	&	0.001903 \\
23	&	martyrs and resistance	&	1877	&	0.001877 \\
24	&	faith in God	&	1853	&	0.001853 \\
25	&	fires incidents	&	1829	&	0.001829 \\
26	&	gaza prayers support	&	1607	&	0.001607 \\
27	&	armed conflicts &	1439	&	0.0014 \\
28	&	moon and space	&	1424	&	0.001424 \\
29	&	weather forecast	&	1385	&	0.001385 \\
30	&	job openings saudiarabia	&	1362	&	0.001362 \\
31	&	seeking knowledge	&	1353	&	0.001353 \\
32	&	gaza airstrikes victims	&	1340	&	0.00134 \\
33	&	remembrances of god	&	1318	&	0.001318 \\
34	&	apps and updates	&	1310	&	0.00131 \\
35	&	account management	&	1277	&	0.001277 \\
36	&	prisoner exchange	&	1251	&	0.001251 \\
37	&	seeking forgiveness	&	1186	&	0.001186 \\
38	&	sudan political developments	&	1085	&	0.001085 \\
        \bottomrule
    \end{tabularx}
    \caption{\captionbf{Top-40 Arabic topics}}
    \label{tab:topics_ar}
\end{table*}

\begin{table*}[h]
    \centering
    \begin{tabularx}{\textwidth}{l l l l}
        \toprule
        \textbf{Topic id} & \textbf{Topic} & \textbf{Message count} & \textbf{Fraction} \\
        \midrule
-1	&	political conspiracy	&	644734	&	0.644734 \\
0	&	football leagues updates	&	11075	&	0.011075 \\
1	&	vaccine debate	&	10577	&	0.010577 \\
2	&	taiwan china relations	&	7139	&	0.007139 \\
3	&	christian faith	&	5121	&	0.005121 \\
4	&	negative perceptions actor	&	4718	&	0.004718 \\
5	&	trudeau protests canada	&	4676	&	0.004676 \\
6	&	election fraud allegations	&	4646	&	0.004646 \\
7	&	aviation incidents	&	4098	&	0.004098 \\
8	&	trump rallies events	&	4061	&	0.004061 \\
9	&	vaccine documentary watching	&	3688	&	0.003688 \\
10	&	islamic teachings	&	3610	&	0.00361 \\
11	&	biden presidency debate	&	3597	&	0.003597 \\
12	&	female characters analysis	&	3227	&	0.003227 \\
13	&	racial discourse	&	3204	&	0.003204 \\
14	&	elon musk twitter	&	2868	&	0.002868 \\
15	&	wildfires and firefighters	&	2591	&	0.002591 \\
16	&	transgender issues	&	2513	&	0.002513 \\
17	&	brazilian political crisis	&	2333	&	0.002333 \\
18	&	epstein maxwell scandal	&	2206	&	0.002206 \\
19	&	australia freedom movement	&	2202	&	0.002202 \\
21	&	nasa spacex missions	&	1909	&	0.001909 \\
20	&	friends with affection	&	1909	&	0.001909 \\
22	&	antisemitic narratives	&	1888	&	0.001888 \\
23	&	inflation rates policies	&	1749	&	0.001749 \\
24	&	nato news updates	&	1729	&	0.001729 \\
25	&	music expressions	&	1709	&	0.001709 \\
26	&	france political protests	&	1661	&	0.001661 \\
27	&	game releases indie	&	1590	&	0.00159 \\
28	&	mass shootings	&	1570	&	0.00157 \\
29	&	samsung smartphones android	&	1568	&	0.001568 \\
30	&	pope francis death	&	1505	&	0.001505 \\
31	&	drone warfare operations	&	1489	&	0.001489 \\
32	&	anthro canine nsfw	&	1469	&	0.001469 \\
33	&	5g emf radiation	&	1398	&	0.001398 \\
34	&	climate change hoax	&	1372	&	0.001372 \\
35	&	abortion debate	&	1347	&	0.001347 \\
36	&	bjp elections india	&	1342	&	0.001342 \\
37	&	jfk assassination conspiracy	&	1336	&	0.001336 \\
38	&	royal family news	&	1323	&	0.001323 \\
        \bottomrule
    \end{tabularx}
    \caption{\captionbf{Top-40 English topics}}
    \label{tab:topics_en}
\end{table*}

\FloatBarrier

\renewcommand{\arraystretch}{1.1}

\begin{table*}[h]
    \centering
     {\footnotesize
    \begin{tabularx}{\textwidth}{ p{0.2\textwidth} p{0.15\textwidth} X}
        \toprule
\textbf{Chat name} & \textbf{Language} & \textbf{Characteristic topic} \\
\midrule
INSIDERR\_POLITIC & English & brics expansion 2023 \\
vanek\_nikolaev & Russian  & military air operations \\
INSIDER\_USA\_NEWS & English & child protection legislation \\
vv\_volodin & Russian  & ukraine crisis mentions \\
Russica2 & Russian  & russia geopolitics conflict \\
dmitrynikotin & Russian  & ukraine celebration media \\
real\_DonaldJTrump & English & election fraud allegations \\
trump\_magacommunity & English & justice fight victims \\
zarubinreporter & Russian  & news updates telegram \\
vatnoeboloto & Russian  & poverty in russia \\
panchenkodi & Russian  & ukraine crisis mentions \\
stalin\_gulag & Russian  & poverty in russia \\
slvn\_pomet & Russian  & ukraine conflict updates \\
project\_veritas & English & undercover investigation \\
PatriaDigital & Spanish  & brics expansion 2023 \\
realKarliBonne & English & trump debate topics \\
JamesOKeefeIII & English & election fraud allegations \\
Alertas24 & Spanish  & protests in Venezuela \\
realx22report & English & congressional events \\
PepeMatter & English & child protection legislation \\
qthestormrider777 & English & deep state exposure \\
CharlieKirk & English & gop election audit \\
RealGenFlynn & English & veteran honoring events \\
LauraAbolichannel & English & stock market crisis \\
TrumpChannel & English & biden hunter laptop \\
stewpeters & English & military degeneracy issues \\
RealDonaldoTrumpo & English & political resignation votes \\
BellumActaNews & English & protests in Venezuela \\
ResisttheMainstream & English & biden trump transition \\
ShadowofEzra & English & twitter censorship health \\
rattletrap1776 & English & space force initiatives \\
DirtRoadDiscussion & English & satanic rituals war \\
TuckerFans & English & trump election interference \\
sanidadgob & Spanish  & health and wellbeing \\
DDGeopolitics & English & ukraine conflict updates \\
bioclandestine & English & freedom fight hope \\
Jack\_Posobiec & English & sexual abuse cases \\
SantaSurfing & English & taliban takeover afghanistan \\
trottasilvano & French  & green energy sustainability \\
DBongino & English & dan bongino streaming \\
        \bottomrule
    \end{tabularx}
    \vspace{0.5em}
    }
\end{table*}

\begin{table*}[h]
    \centering
     {\footnotesize
    \begin{tabularx}{\textwidth}{ p{0.2\textwidth} p{0.15\textwidth} X}
        \toprule
\textbf{Chat name} & \textbf{Language} & \textbf{Characteristic topic} \\
\midrule
liusivaya & Spanish  & ukraine crisis mentions \\
Richardcitizenjournalists & English & news updates telegram \\
VigilantFox & English & meat health supplements \\
rsbnetwork & English & trump debate topics \\
FearlessReport & English & biden hunter laptop \\
AntiSpiegel & German & eu regulations analysis \\
SGTnewsNetwork & English & child protection legislation \\
JFK\_Q17 & English & truth unleashed \\
geopolitics\_live & English & ukraine air defense \\
CaptKylePatriots & English & elite crime exposures \\
cruel\_historyy & English & iraq protests violence \\
Slavyangrad & English & artillery operations \\
DanScavinoFORCE & English & election fraud allegations \\
HATSTRUTH & English & spiritual struggles prayer \\
ReinerFuellmichEnglish & English & twitter censorship health \\
andweknowLT & English & disinformation alerts \\
MichaelJLindell & English & election fraud allegations \\
IntelRepublic & English & russia european relations \\
candlesinthenight & English & citizenship exchange service \\
QNewsOfficialTV & Dutch & meat health supplements \\
FaktenFriedenFreiheit & German   & medical controls criticism \\
GitmoTV & English & elite crime exposures \\
ScottRitter & English & citizenship exchange service \\
JamesWoodsFans & English & biden trump transition \\
worlddoctorsalliance & English & olivia health issues \\
stormypatriotjoe21 & English & luciferian satanism \\
projectcamelotKerry & English & military intelligence news \\
patriotstreetfighter & English & event tour schedule \\
NewsmaxTV & English & newsmax podcasts \\
followsthewhiterabbit & English & military air operations \\
ElectionHQ2024 & English & election fraud allegations \\
RealMarjorieGreene & English & blm antifa violence \\
AMGNEWS2022 & English & elite crime exposures \\
GarrettZ & English & election fraud cronyism \\
Qanon\_storm\_incoming & English & spiritual struggles prayer \\
robinmg & English & eu regulations analysis \\
BennyJohnson & English & biden trump transition \\
police\_frequency & English & protests in Venezuela \\
LizCrokinReport & English & epstein scandal \\
PrincessDiana\_Q & English & biden trump transition \\
        \bottomrule
    \end{tabularx}
    \vspace{0.5em}
    }
\end{table*}

\begin{table*}[t]
    \centering
     {\footnotesize
    \begin{tabularx}{\textwidth}{ p{0.2\textwidth} p{0.15\textwidth} X}
        \toprule
\textbf{Chat name} & \textbf{Language} & \textbf{Characteristic topic} \\
\midrule
TheTrumpist & English & lebron james crash \\
BenShapiroGang & English & congressional events \\
Haintz & German   & health and wellbeing \\
ConservativeBrief & English & biden trump transition \\
DonaldTrumpOffice & English & multilateral dialogue \\
NTDNews & English & green energy sustainability \\
hsretoucher17 & English & trump debate topics \\
zeeemedia & English & injection injuries athletes \\
TrueGreatAwakening & English & secret control influence \\
BannonWarRoom & English & eu regulations analysis \\
WesternJournal & English & lebron james crash \\
techno\_fog & English & epstein scandal \\
liltalkshow & English & whistleblower dominion bios \\
TheRealKimShady7 & English & luciferian satanism \\
PepeDeluxed & English & democracy peril corporation \\
theprofessorsrecord & English & election fraud allegations \\
WhiteHatsQ & English & deep state exposure \\
wendyrogersaz & English & senate legislation minors \\
DrDavidMartin & English & democracy peril corporation \\
jordansather & English & content moderation \\
        \bottomrule
    \end{tabularx}
    \vspace{0.5em}
    }
    \caption{\captionbf{Initial seed list of Telegram channels used to bootstrap the snowball crawler.} We picked the 100 biggest channels in the political category according to \url{https://telegramchannels.me/ranking} (retrieved on 2025-05-15). The characteristic topic is the topic that is most overrepresented in messages from the given chat, compared to the baseline of messages from all seed chats.}
    \label{tab:seed_list}
\end{table*}

\begin{algorithm*}[h]
\caption{Snowball crawling of Telegram channels. The algorithm begins with an initial channel, extracts forwarded messages to discover new channels, and repeats until no new channels are found. Distribution across multiple workers and queue prioritization by out-degree are omitted for clarity.}
\label{alg:snowball_crawling}
\hrulefill \\
\textbf{Input}: Initial channel $C_{\text{initial}}$ \\
\textbf{Output}: List of discovered Telegram channels
\begin{algorithmic}[1]
\STATE $q \gets \{C_{\text{initial}}\}$
\WHILE{$q$ is not empty}
    \STATE $C_{\text{current}} \gets \text{Dequeue}(q)$
    \STATE $m \gets \text{GetMessages}(C_{\text{current}})$
    \STATE $m_f \gets \text{GetForwardedMessages}(m)$
    \STATE $C_{\text{new}} \gets \text{DiscoverChannels}(m_f)$
    \STATE $\text{Enqueue}(q, C_{\text{new}})$
\ENDWHILE
\end{algorithmic}
\end{algorithm*}

\end{document}